\documentstyle[preprint,eqsecnum,aps,epsfig]{revtex}

\begin{document}
\draft

\title{Limits on the cosmological abundance of supermassive compact objects from a millilensing search in gamma-ray burst data}

\author{Robert J. Nemiroff}
\address{Michigan Technological University, Department of Physics,
1400 Townsend Drive, Houghton, MI  49931}

\author{Gabriela F. Marani\cite{NRC}, Jay P. Norris, and 
Jerry T. Bonnell\cite{USRA}}
\address{NASA Goddard Space Flight Center, Code 661, 
Greenbelt, MD  20771}

\date{\today}
\maketitle

\begin{abstract}
A new search for the gravitational lens effects of a significant cosmological density of supermassive compact objects (SCOs) on gamma-ray bursts has yielded a null 
result.  We inspected the timing data of 774 BATSE-triggered GRBs for evidence of millilensing: repeated peaks similar in light-curve shape and spectra.  Our null 
detection leads us to conclude that, in all candidate universes simulated, $\Omega_{SCO} < 0.1$ is favored for $10^5 < M_{SCO}/M_{\odot} < 10^9$, while in some universes and mass ranges the density limits are as much as 10 times 
lower. Therefore, a cosmologically significant population of SCOs near globular cluster mass neither came out of the primordial universe, nor condensed at recombination.
\end{abstract}


\narrowtext

\section{Introduction}
\label{Introduction}

As gamma ray bursts (GRBs) occur in the distant universe, their light is susceptible to gravitational lensing by intervening matter.  In the gamma-ray band, the spatial 
resolution of current detectors is not great enough to allow the detection of individual GRB images, but the microsecond temporal resolution of many GRB detectors might 
allow for the detection of individual images temporally \cite{Nem98}.  

The nature of much of the dark matter in the universe remains unknown.  Carr \& Rees \cite{Car84} and Gnedin \& Ostriker \cite{Gne92} suggested the theoretical possibility that condensates near the Jean's mass 10$^{6.5} M_{\odot}$ could have been the first pregalactic objects, created near the time of recombination in the early universe, with their mass and cosmological density depending on conditions.  On 
observational grounds, globular clusters are the oldest objects known, and darker, more compact objects on their mass scale might well have also been created and escaped detection \cite{Car94} \cite{Car99}. We will refer to these dark matter candidates as Supermassive Compact Objects (SCOs).  

Paczynski \cite{Pac87} first suggested the possibility that GRBs could be seen to undergo gravitational lensing. The very short durations of some GRBs make them good sources for the lens detections of SCOs \cite{Bla92}, a type of lensing known as millilensing.   Previously, Nemiroff et al. \cite{Nem93} searched 44 GRBs for millilensing, and the null detection was used to show the universe was not filled to closure density of such lenses \cite{Nem93}. Using the same data, the ($\Omega_M$, $\Omega_{\Lambda}$) $=$ (0.15, 0.0) cosmology \cite{Gne92} was marginally ruled out Marani et al. \cite{Mar99a}.

To act as a gravitational lens, a SCO must be compact enough to avoid absorbing one of the passing images.  Crudely, this means that the lens must be smaller than its own Einstein ring.  For the above canonical mass at a redshift of unity in a smooth ($\Omega_M$, $\Omega_{\Lambda}$) $=$ (0.70, 0.30), this radius is about 25 pc.  Some massive, compact globular clusters could make good millilenses, but many others would not.  Black holes, of course, fit all SCO and millilens requirements.

In Section 2, we report the details of a new search for millilensing on 774 GRBs, a factor of $>$15 increase in sources over the last published millilensing search 
\cite{Nem93}. Again, no millilensing candidates were found. In Section 3, we describe the theoretical implications of this null detection in terms of limits on cosmological abundances of SCOs.  In Section 4, we provide some discussions and conclusions.

\section{The Millilensing Search}

Since gravitation does not create a path-dependant time-dilation effect on passing photons, the light-curves of different gravitational lens-induced GRB images are expected to be identical.  Similarly, since gravitation deflects photons of different energy equally, the spectra of different gravitational lens-induced GRB images are expected to be identical.  We therefore search for millilensing by comparing the spectra and the light curves of different peaks in the time series of GRBs.  The expected time delay between GRB images created by a single SCO is \cite{Kra91} 
\cite{Mar99a}
 \begin{equation}
 \Delta t = (1+z_l) \left({(1-f) \over \sqrt{f}} + {\rm ln} f \right) R_S/c
 \end{equation}
where $z_l$ is the redshift of the SCO lens, $f$ is the ratio in brightness between the brightest two images and is defined to be less than unity, $c$ is the speed of light, 
$R_S$ is the Schwarzschild radius of the SCO lens, and $M$ is the mass of the SCO lens.  Note that for canonical values like $M = 10^7 M_{\odot}$, $f = 0.3$, and $z_l=1.5$, the expected time delay between images would be on the order of 20 seconds.

The GRBs we inspected for millilensing were detected by the Burst and Transient Source Experiment (BATSE) onboard the Compton Gamma Ray Observatory (CGRO).   This search was the first fully automated search.  GRB data were only inspected visually if they passed several automated millilensing tests.  Initial candidates included all GRBs detected between 1991 April and 1999 January that had 64-ms ASCII data available on-line at the Compton Gamma Ray Observatory 
Science Support Center: 1796 GRBs in all.  The last GRB investigated was BATSE trigger number 7353.  The data typically spanned about up to 524 seconds. 

GRBs were eliminated from the candidate millilensing sample if they had any data gaps.  GRBs with data gaps would likely not show any type of millilensing bias but would pose special challenges for a program written in a general, automated fashion.  Surviving GRBs were then eliminated from the candidate millilensing sample if they had backgrounds that sloped by more than 4.5 $\sigma$ between the first 64-ms time bin and the last 64-ms time bin.  Again, these GRBs should show little bias toward millilensing but posed unwieldy programming challenges.  After these cuts 774 GRB millilensing candidates survived.

The time series of all surviving GRBs were then analyzed by an automated millilensing search program: millisearch.for.  The program's first task was to isolate the first discrete episode of GRB emission, which we will call the ``burst."  To start, the highest 64-ms time bin of each surviving GRB was first found.  This high bin was then added to the very next trailing bin to see if the signal to noise ($S/N$) above background was increased.  If not, the burst was considered to have ended.  If so, the trailing bin was considered part of the burst, and added to the burst.  The next trailing bin was then considered and treated identically.  After the last trailing bin was added to the burst, leading bins were then tested and added, again until the $S/N$ over background was found to decrease.  At this point, the first contiguous episode of GRB emission was taken to be defined.

Next, the program searched to see if a second identical (to within an intensity scale factor), gravitational-lens induced image of the burst could possibly have followed in the available BATSE time series.  A candidate emission episode is referred to here as an ``echo."  To find an echo, the highest 64-ms time-bin following at least one full burst duration after the conclusion of the burst was first isolated. A series of 64-ms time bins around the echo peak was then isolated equal to the duration of the burst.  
GRBs where the echo was below 5.5 $\sigma$ over the background were recorded as having no significant echo.  For those GRBs without a candidate echo, maximum echo strength $f_{max}$, minimum time delay between burst and echo $t_{min}$, and maximum time delay between burst and echo $t_{max}$ were recorded for the analysis described in the next section.

Those GRBs with a candidate echo had $f_{max}$ recorded as the counts ratio between the echo and the burst.  As before $t_{min}$ and $t_{max}$ were also recorded for these GRBs.  All the counts in all the time bins of the burst and echo were then co-added in all four major BATSE energy channels.  The relative counts in each energy channel were compared between burst and echo by a two-distribution $\chi^2$ test.  Eleven GRBs passed this test, and were then compared for light-curve similarity.  One such candidate is shown in Figure 1.  No GRBs were found to have both matching millilensing spectra and light-curves.  The final output of the millilensing search program was a list of BATSE trigger numbers, and $f_{max}$, $t_{min}$, and $t_{max}$ for each trigger.  

Faux GRBs with a fictitious millilensing signal were used to successfully test the millilensing detection program.  This program, millisearch.for, is available through 
the Astrophysics Source Code Library at http://ascl.net.

\section{Cosmogonic Implications}

As millilensing would be expected in a variety of cosmogonic scenarios, non-detection rules out some of these scenarios, and sets limits on the potential abundance of SCOs in other scenarios.  To compute these limits, we used the detection volume formalism \cite{Nem89} \cite{Nem93}.   To be detectable, lenses must fall inside a volume where they would create two bright images with dynamic range between them $f_{min}$.  Additionally, lenses must fall inside another volume where the two bright images would be separated in time by less than a maximum amount $t_{max}$, and fall {\it outside} a third volume where the two bright images would be separated in time by less than a minimum amount $t_{min}$. 

At each lens redshift $z_l$, a detection volume has a radius $b$.  For the detection criterion of magnification, the detection volume radius is given in \cite{Nem89} as 
$b_A = \sqrt{4 R_S D^A_{OL} D^A_{LS} \Phi / D^A_{OS}}$ where $R_S$ is the Schwarzschild radius of the compact lens, $D^A$ refers to angular diameter distance, and subscripts $O, L, \& S$ refer to the observer, lens and source respectively.  Magnification $A$ is embedded in $\Phi = A / \sqrt{A^2 - 1} - 1$.  Dynamic range $f$ between the two brightest images of a compact lens is by definition less than unity and related to total magnification by $A = (1 + f) / (1 - f)$.  Similarly, the time delay $\Delta t$ between the two brightest images of a compact lens is related to dynamic range by Eqn. 2.1.
 
A lens is detectable if it falls inside both $b_f$ and $b_{tmax}$ but outside $b_{tmin}$.  The total number of detectable millilenses is therefore given by 
 \begin{equation}
 N_{lens} = \int_{0}^{z_s} n_l \pi  (b_{outer}^2 - b_{inner}^2) dD_{OL}^P
 \end{equation}
where $D_{OL}^P$ is the proper distance between the observer and the lens, $n_l$ is the comoving lens density, $b_{outer}$ is the radius of the outermost detection 
volume where $b_{outer} =$ min $(b_f, b_{tmax})$ and $b_{inner} = b_{tmin}$.  

In a few cases, GRB source redshifts $z_s$ have been actually determined (see, for example, \cite{Met97}).  To find the likely effects of millilensing, however, we needed redshifts for the whole BATSE sample.  We therefore estimated these redshifts from Monte Carlo simulations.  These simulations are described in detail elsewhere \cite{Mar99b} \cite{Nem00a}, and have the following characteristics.  GRBs are thrown randomly in a universe with a specified geometry.  These GRBs are assigned a generic spectrum and a peak flux chosen from a published luminosity function \cite{Kom97}.  These GRBs are then redshifted and re-sampled to see if they would re-trigger BATSE.  Those that do are fitted to the BATSE 3B detection 
rate, the BATSE Log $N$ - Log $P$, time-dilation \cite{Nor94} \cite{Bon97}, and $\langle V/V_{max} \rangle$.  The later statistic supplemented the normalization of the Log $N$ - Log $P$ fit, which was most sensitive to brightness distribution shape.  Adjustable free parameters included an integrated flux scale factor $L_o$ and a scale factor on the comoving number density $n_s$.  A file recording acceptable fits for each universe geometry was recorded.  Each of the 774 BATSE millilensing candidates was then matched-up to the Monte-Carlo GRB with the closest peak flux, and assigned the redshift of this Monte-Carlo GRB.

Results are shown for two of these cosmologies in Figures 2 and 3.  In the figures, the number of expected lens detections is plotted as a function of lens mass and density.  The contours represent 1, 5, and 10 expected lens 
detections, and hence roughly correspond to exclusions at the 1, 2, and 3 $\sigma$ levels.    

The expected number of lens detections in a ($\Omega_M$, $\Omega_{\Lambda}$) $=$ (0.3, 0.7) universe \cite{Rei98} \cite{Per99} is shown in Figure 2, on the left.  $\Omega_M$ is the density of matter in the universe, relative to the critical density, while $\Omega_{\Lambda}$ is the density in a cosmological constant.

Here one finds a wide range of masses is excluded for several interesting values of $\Omega_{CO}$.  In fact, $\Omega_{CO}$ dips below even 0.03 for a narrow range 
of masses near $M/M_{\odot} \sim 10^7$.  Therefore, even a significant amount of baryonic dark matter is excluded from forming compact lenses in this mass range.

To test Gnedin \& Ostriker's \cite{Gne92} universe directly, Figure 3, on the right, shows similar results for a (0.15, 0) universe.  It appears difficult for any 
significant fraction of the mass density to hide in supermassive compact objects in the mass range $10^6 < M/M_{\odot} < 10^8$.

\section{Discussion and Conclusions}

Lensing of GRBs has the ability to time-resolve a mass range of dark matter not usually probed: that of compact objects near globular cluster mass.  That these lenses 
were not detected indicates, as we have shown above, that the majority of the universe's mass density does not reside in SCOs.  

A potential inaccuracy in the above analysis lies in the estimated GRB redshifts.  The average GRB redshift we have estimated is between 1.5 and 2, like most so far determined in the optical.  However, some popular speculation holds that average BATSE redshifts would be higher, and, if true, would make the limits estimated here even more conservative.  

The above lens rates did not incorporate the potential effects of magnification bias.  Such bias \cite{Bar99} would be expected were the GRB luminosity function to rise sharply near the effective redshift limit of the BATSE sample.  One reason for not including magnification bias is the potential uncertainty in the slope of the GRB luminosity function.  As including magnification bias might significantly increase the expected amount of lensing, it makes the estimated limits derived here even more conservative.

Another potential inaccuracy would occur were different GRBs images to have different light-curves or spectra.  This might occur were GRBs to undergo beaming 
\cite{Fru99,Wij99,Har99} on the order of milli-arcseconds or less, something that has not been ruled out by observation. 

As more GRBs become available for millilensing inspection, the ability to find a smaller and smaller abundance of SCOs becomes possible.  Potentially, as much as a factor of 10 more GRBs might become available, allowing an abundance as low as $\Omega_{SCO} \sim 0.003$ to be discernable.  A survey sensitive to more GRBs and greater time resolution might detect the known star-field in GRB lensing, as well 
as dark matter on scales near one solar mass \cite{Nem98}.

\acknowledgments

This research was supported, in part, by grants by NASA.  RJN acknowledges additional support from a grant by the NSF.  GFM acknowledges additional support from the NRC. We thank Chip Meegan for helpful discussions.

Note added in proof: After most of this work was completed,
we became aware of work by Wilkinson and collaborators who were
making a similar investigation into cosmological millilensing
but using observations with VLBA radio telescopes.  That group
reports their work and results in the next paper.

\begin{figure}
\epsfig{file=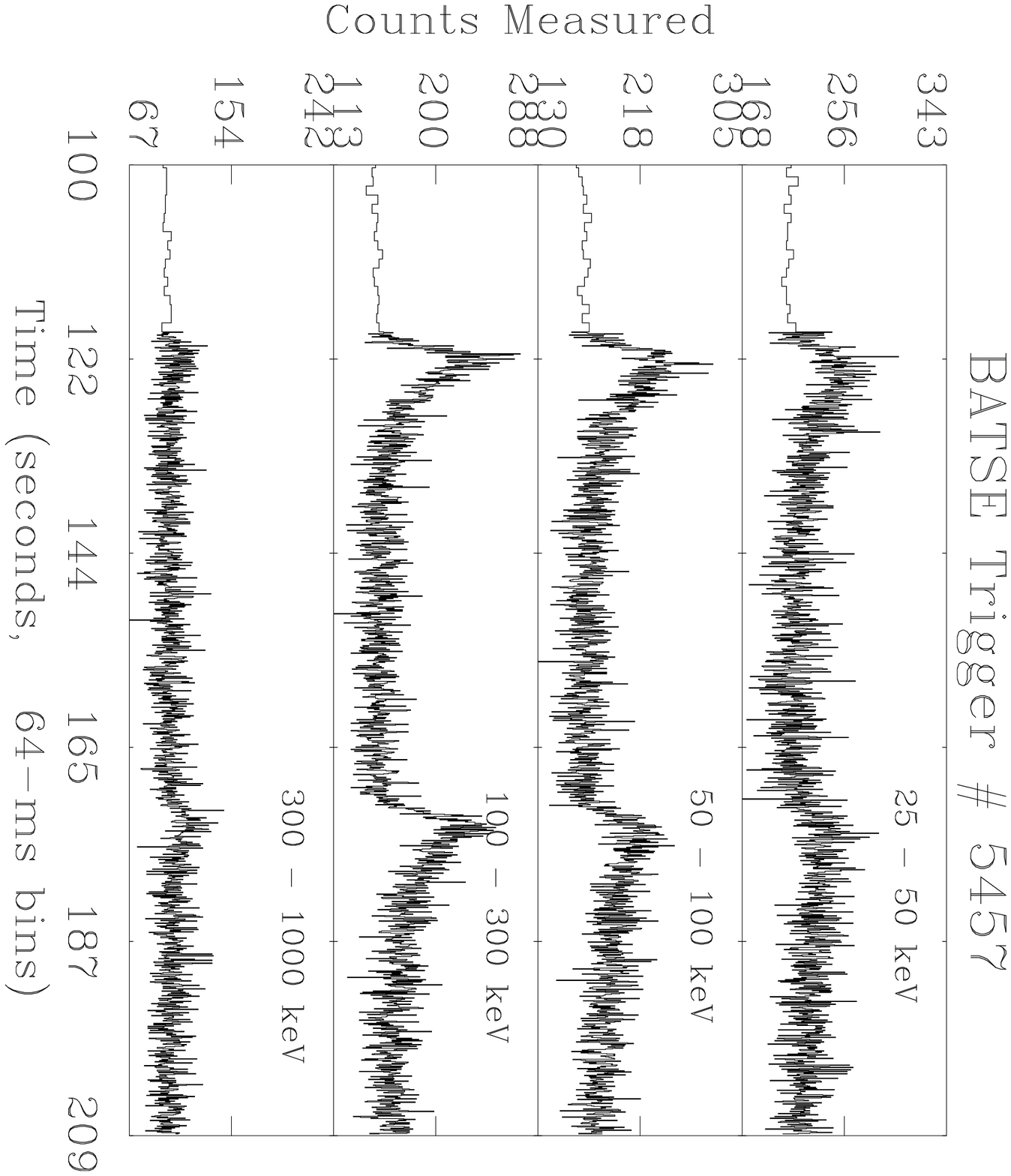}
\caption{BATSE GRB trigger 5457 light-curve showing two emission episodes that comprised one of the better candidates for millilensing.  The counts in four energy channels are shown as a function of time.  The time resolution of the data and plot is 64-ms, except before trigger, where it is 1024-ms.}
\end{figure}

\begin{figure}
\epsfig{file=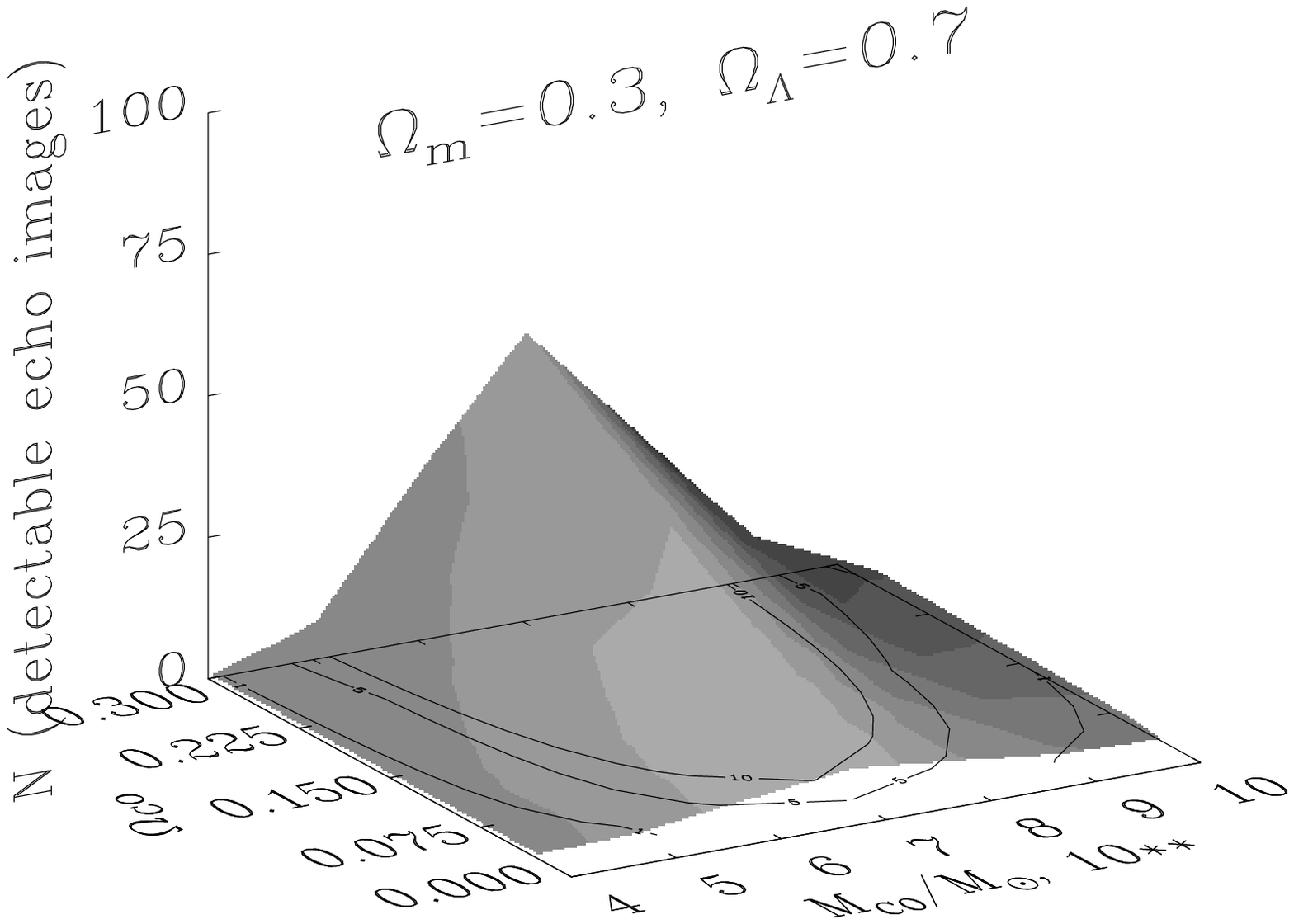}
\caption{The number of expected millilenses in a (0.3, 0.7) universe as a function of millilens mass, density, and universe geometry.}
\end{figure}

\begin{figure}
\epsfig{file=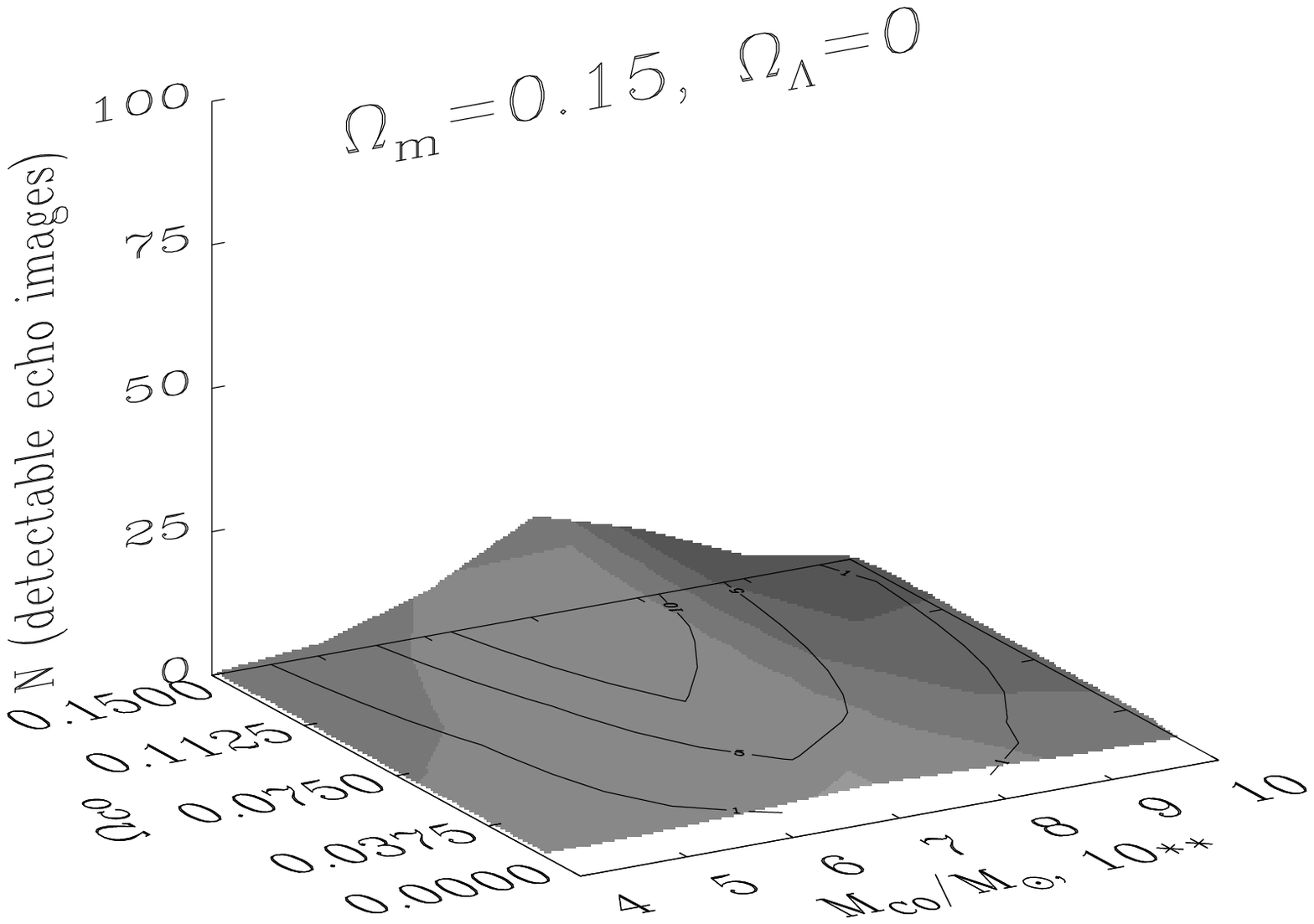}
\caption{The number of expected millilenses in a (0.15, 0) universe as a function of millilens mass, density, and universe geometry.}
\end{figure}

\end{document}